\documentclass[conference,a4paper]{APSIPA2019}
\usepackage{multirow}
\usepackage[pdftex]{graphicx}
\usepackage{amsmath}
\usepackage[psamsfonts]{amssymb}
\usepackage{amsxtra}
\usepackage{threeparttable}
\usepackage{cite}

\usepackage{bm}
\usepackage[linesnumbered,lined,ruled]{algorithm2e}

\newcommand{\bhline}[1]{\noalign{\hrule height #1}}

\newcommand{\tr}{\mathrm{tr}}
\newcommand{\rbm}[1]{\bm{\mathrm{#1}}}
\newcommand{\HH}{\mathsf{H}}
\renewcommand{\top}{\mathsf{T}}
\renewcommand{\Re}{\mathrm{Re}}

\newcommand{\xij}{\bm{x}_{ij}}
\newcommand{\hij}{\bm{h}_{ij}}
\newcommand{\uij}{\bm{u}_{ij}}
\newcommand{\sij}{s_{ij}^{(h)}}
\newcommand{\ai}{\bm{a}_{i}^{(h)}}
\newcommand{\bi}{\bm{b}_{i}}
\newcommand{\vpi}{\bm{b}_{i}}

\newcommand{\rh}{r_{ij}^{(h)}}
\newcommand{\ru}{r_{ij}^{(u)}}
\newcommand{\li}{\lambda_{i}}
\newcommand{\ri}{\rbm{R}_{i}^{(u)}}
\newcommand{\rit}{\rbm{R}_{i}'^{(u)}}

\newcommand{\hrh}{\hat{r}_{ij}^{(h)}}
\newcommand{\hru}{\hat{\rbm{R}}_{ij}^{(u)}}

\newcommand{\trh}{\tilde{r}_{ij}^{(h)}}
\newcommand{\tru}{\tilde{r}_{ij}^{(u)}}
\newcommand{\tli}{\tilde{\lambda}_{i}}
\newcommand{\tri}{\tilde{\rbm{R}}_{i}^{(u)}}
\newcommand{\trijx}{\tilde{\rbm{R}}_{ij}^{(x)}}

\newcommand{\tal}{\tilde{\gamma}_{ij}}

\newcommand{\raa}{\rho_{i}^{(aa)}}
\newcommand{\rax}{\rho_{ij}^{(ax)}}
\newcommand{\rxx}{\rho_{ij}^{(xx)}}
\newcommand{\traa}{\tilde{\rho}_{i}^{(aa)}}
\newcommand{\trax}{\tilde{\rho}_{ij}^{(ax)}}
\newcommand{\sap}{\sigma_{i}^{(ab)}}
\newcommand{\spx}{\sigma_{ij}^{(bx)}}
\newcommand{\taa}{\tau_{i}^{(aa)}}
\newcommand{\tax}{\tau_{ij}^{(ax)}}
\newcommand{\txx}{\tau_{ij}^{(xx)}}

\begin{document}
\allowdisplaybreaks[1]

\title{Acceleration of rank-constrained spatial covariance matrix estimation for blind speech extraction}

\author{%
\authorblockN{%
Yuki Kubo\authorrefmark{1},
Norihiro Takamune\authorrefmark{1},
Daichi Kitamura\authorrefmark{2}, and
Hiroshi Saruwatari\authorrefmark{1}
}
\authorblockA{%
\authorrefmark{1}
The University of Tokyo, Graduate School of Information Science and Technology, Tokyo, Japan}
\authorblockA{%
\authorrefmark{2}
National Institute of Technology, Kagawa College, Kagawa, Japan}
}

\maketitle
\thispagestyle{empty}

\begin{abstract}
    In this paper, we propose new accelerated update rules for rank-constrained spatial covariance model estimation, 
    which efficiently extracts a directional target source in diffuse background noise.
    The naive update rule requires heavy computation such as matrix inversion or matrix multiplication.
    We resolve this problem by expanding matrix inversion to reduce computational complexity;
    in the parameter update step, we need neither matrix inversion nor multiplication.
    In an experiment, we show that the proposed accelerated update rule achieves 87 times faster
    calculation than the naive one.
\end{abstract}

\section{Introduction}
Blind source separation (BSS)~\cite{Sawada2019_BSS} is a technique that separates an observed multichannel signal
into each source signal without any prior information about each source or the mixing system.
In a determined or overdetermined situation (number of sensors $\geq$ number of sources),
frequency-domain independent component analysis (FDICA)~\cite{PSmaragdis1998_BSS,Araki2003_FDICAlim,HSaruwatari2006_FDICA},
independent vector analysis (IVA)~\cite{AHiroe2006_IVA,TKim2007_IVA,NOno2011_AuxIVA},
and independent low-rank matrix analysis (ILRMA)~\cite{DKitamura2016_ILRMA,DKitamura2018_ILRMA,Kitamura2018_tGGDILRMA,Mogami2018_GGDILRMA,Mogami2019_GKLILRMA}
have been proposed for audio BSS problems.

In this paper, we address a situation where a directional target source and diffuse noise that arrives from all directions are mixed.
In this case, FDICA, IVA, and ILRMA do not enable the extraction of only the target source in principle~\cite{SAraki2003_ICAeqBF}, and the estimated target source includes residual diffuse noise.

To model such diffuse (spatially spread) noise,
a multichannel extension of nonnegative matrix factorization
(multichannel NMF: MNMF)~\cite{AOzerov2010_MNMF,HSawada2013_MNMF,Nikunen2014_DOASCM}
was proposed.
MNMF estimates a full-rank spatial covariance matrix (SCM)~\cite{Duong2010_SpatialMatrix},
which represents the time-invariant spatial characteristics of each source.
However, since the number of parameters in MNMF is large, its optimization requires a huge computational cost and
lacks robustness against the initialization~\cite{DKitamura2016_ILRMA}.

FastMNMF~\cite{Ito2019_FastMNMF, Sekiguchi2019_FastMNMF} enables computationally efficient estimation
by introducing a jointly diagonalizable SCM into the MNMF model.
Although this assumption greatly reduces the computational cost of the update algorithm,
it still suffers from dependence on the initial values.

To solve the above-mentioned problems,
we proposed rank-constrained SCM estimation~\cite{Kubo2019_RCSCM}, which is a postprocessing method for ILRMA.
In this method, to effectively model diffuse noise in an overdetermined situation,
a full-rank SCM of diffuse noise is recovered from the estimates obtained by ILRMA.
Since ILRMA can precisely cancel the directional target source~\cite{YTakahashi2009_BSSA},
an accurate rank-$(M\!-\!1)$ SCM of diffuse noise can be obtained.
Thus, the rank-constrained SCM estimation enables us to restore the lost spatial basis of diffuse noise,
resulting in accurate extraction of the directional target source.
By employing ILRMA as a preprocessing method, we can achieve both lower-cost computation
and more initialization-robust estimation compared with MNMFs.

In this paper, we present a more computationally efficient algorithm
for rank-constrained SCM estimation.
The naive update rule in rank-constrained SCM estimation
(hereafter referred to as \textit{naive update})
requires matrix inversion at each time-frequency slot,
which leads to a heavy computational load.
To solve this problem, we propose an accelerated algorithm
that expands the inversion of matrices using the Sherman--Morrison formula
and pseudoinverse.
The computational cost of each update is reduced from $O(IJM^3)$
to $O(IJ)$, where $I$ and $J$ are the numbers of frequency bins
and time frames, respectively.
The efficacy of the proposed algorithm is confirmed via BSS
with directional target speech and diffuse babble noise.

\section{Rank-Constrained SCM Estimation}
\subsection{Formulation}
Let us denote a source signal and a multichannel observed signal as $\bm{s}_{ij}=(s_{ij,1},\dots,s_{ij,n},\dots,s_{ij,N})^{\top}\in\mathbb{C}^{N}$
and $\xij=(x_{ij,1},\dots,x_{ij,m},\dots,x_{ij,M})^{\top}\in\mathbb{C}^{M}$,
where $i=1,\dots,I$, $j=1,\dots,J$, $n=1,\dots,N$, and $m=1,\dots,M$ are the indices of the
frequency bins, time frames, microphones, and sources, respectively,
$N$ is the number of sources, and $^{\top}$ denotes the transpose.

If each source can be considered a point source and the reverberation time is sufficiently shorter
than the window length in a short-time Fourier transform (STFT), there exists a mixing matrix
$\rbm{A}_{i}=\left(\bm{a}_{i,1}\cdots\bm{a}_{i,N}\right)\in\mathbb{C}^{M\times N}$ for each frequency bin and the following holds:
\begin{align}
    \xij=\rbm{A}_{i}\bm{s}_{ij},
\end{align}
where $\bm{a}_{i,n}$ is the steering vector of source $n$ at frequency $i$.

When $M \geq N$, independence-based BSS, such as FDICA, IVA, and ILRMA,
can be applied to estimate the demixing matrix $\rbm{W}_{i}=\rbm{A}_{i}^{-1}$,
where dimensionality reduction is used so that $M=N$.
In this work, we use ILRMA as a state-of-the-art BSS method.
For the case of a mixture of one directional target source and diffuse noise,
ILRMA outputs the directional target source estimate and $M-1$ diffuse noise components,
although the directional target source estimate contains 
many diffuse noise components in the same direction.
Therefore, we can calculate $M$ rank-1 SCMs, namely, one rank-1 SCM of the directional target source and
$M-1$ rank-1 SCMs of the diffuse noise components, from the estimated demixing matrix $\rbm{W}_{i}$.
One of these SCMs corresponds to the target source, and the others are components of diffuse noise.
A rank-$(M\!-\!1)$ SCM of diffuse noise can be obtained by summing the $M-1$ rank-1 SCMs.
Since the rank of the SCM of diffuse noise is $M-1$,
we need to restore one lost spatial basis (steering vector) of diffuse noise
to extract the directional target source with a multichannel Wiener filter.

\subsection{Rank-constrained SCM Estimation}
\subsubsection{Generative model}
We assume the observed signal $\xij$ to be the sum of two components, as
\begin{align}
    \xij=\hij+\uij,
\end{align}
where $\hij=(h_{ij,1},\dots,h_{ij,M})^{\top}\in\mathbb{C}^{M}$ and $\uij=(u_{ij,1},\dots,u_{ij,M})^{\top}\in\mathbb{C}^{M}$
are spatial images of the directional target source and diffuse noise, respectively.
$\hij$ is modeled as
\begin{align}
    \hij &= \ai\sij,\\
    \sij|\rh&\sim\mathcal{N}_{c}(0,\rh),\\
    \rh&\sim\mathcal{IG}(\alpha,\beta),
\end{align}
where $\ai$ is the $n_h$th steering vector $\bm{a}_{i,n_h}$, $\sij$ is the dry source of the directional target source $s_{ij,n_h}$,
$\rh$ is the power spectrogram of the $n_h$th source, and $n_h$ denotes the index of the directional target source.
    Here, $\mathcal{N}_{c}$ and $\mathcal{IG}$ denote the circularly symmetric complex Gaussian distribution and the inverse gamma distribution, respectively:
\begin{align}
    \mathcal{N}_{c}(\sij|0,\rh) &= \frac{1}{\pi\rh}\exp\left(-\frac{|\sij|^2}{\rh}\right),\\
    \mathcal{IG}(\rh;\alpha,\beta) &= \frac{\beta^{\alpha}}{\Gamma(\alpha)}(\rh)^{-\alpha-1}\exp\left(-\frac{\beta}{\rh}\right),
\end{align}
where $\alpha>0$ and $\beta>0$ are the shape and scale parameter of the inverse gamma distribution, respectively.
Introducing the above prior distribution improves the estimation performance.

The generative model of the diffuse noise $\uij$ is assumed
to be the multivariate complex Gaussian distribution:
\begin{align}
    \uij &\sim \mathcal{N}_{c}(\bm{0},\ru\ri),\\
    p(\uij) &= \frac{1}{\pi^M(\ru)^{M}|\det\ri|}\exp\left(-\frac{\uij^{\HH}(\ri)^{-1}\uij}{\ru}\right),
\end{align}
where $\ru$ and $\ri$ are the variance and full-rank SCM of diffuse noise, respectively.
In rank-constrained SCM estimation, $\ri$ is expressed using separation filter $\rbm{W}_{i}=(\bm{w}_{i,1}\cdots\bm{w}_{i,M})^{\HH}$ estimated by ILRMA as follows:
\begin{align}
    \ri &= \rit+\li\bi\bi^{\HH},\\
    \rit &= \frac{1}{J}\sum_{j}\rbm{W}_{i}^{-1}(|\bm{w}_{i,1}^{\HH}\xij|^2,\dots,|\bm{w}_{i,n_h-1}^{\HH}\xij|^2,0,\nonumber\\
    &\phantom{=}|\bm{w}_{i,n_h+1}^{\HH}\xij|^2,\dots,|\bm{w}_{i,M}^{\HH}\xij|^2)(\rbm{W}_{i}^{-1})^{\HH}\label{eq:rit},
\end{align}
where $\rit$ is the rank-$(M\!-\!1)$ SCM estimated by ILRMA, 
$\bi$ is the unit eigenvector of $\rit$ that corresponds to the zero eigenvalue,
$\li$ is a scalar variable, and $^{\HH}$ denotes the Hermitian transpose.
Equation (\ref{eq:rit}) applies back-projection~\cite{NMurata2001_Permutation} to $M-1$ diffuse noise components (estimates)
and calculates the sum of the $M-1$ resultant SCMs.
Thus, we fix both the spatial basis of the directional target source $\ai$
and the rank-$(M\!-\!1)$ SCM $\rit$.
The full-rank SCM $\ri$ can be recovered by estimating the residual eigenvalue $\li$.
Also, the variances of the directional target source and diffuse noise, $\rh$ and $\ru$, respectively, are simultaneously estimated.

\subsubsection{Parameter Estimation}
We use the EM algorithm to estimate parameters $\rh$, $\ru$, and $\li$ by maximum a posteriori estimation.
A $Q$ function is defined by the expectation of the complete log-likelihood with respect to the posterior probability of latent variables $\sij$ and $\uij$ as
\begin{align}
    Q(\Theta;\tilde{\Theta}) &= \sum_{i,j}\left[-(\alpha+2)\log\rh-\frac{\hrh+\beta}{\rh}-M\log\ru\right.\nonumber\\
    &\phantom{=}\left.\mbox{}-\log\det\ri-\frac{\tr((\ri)^{-1}\hru)}{\ru}\right]+\mathrm{const.},
\end{align}
where $\mathrm{const.}$ includes the constant terms independent of the parameters,
$\Theta=\{\rh,\ru,\li\}$ is the set of parameters to be updated,
$\tilde{\Theta}=\{\trh,\tru,\tli\}$ is the set of up-to-date parameters,
and $\hrh$ and $\hru$ are the sufficient statistics obtained by the following E-step update rules:
\begin{align}
    \tri &= \rit + \tli\bi\bi^{\HH}\\
    \trijx &= \trh\ai(\ai)^{\HH}+\tru\tri,\label{eq:rijx}\\
    \hrh  &= \trh-(\trh)^2(\ai)^{\HH}(\trijx)^{-1}\ai\nonumber\\
        &\phantom{=} +|\trh\xij^{\HH}(\trijx)^{-1}\ai|^2,\label{eq:E_hrh}\\
    \hru  &= \tru\tri - (\tru)^2\tri(\trijx)^{-1}\tri\nonumber\\
        &\phantom{=} +(\tru)^{2}\tri(\trijx)^{-1}\xij\xij^{\HH}(\trijx)^{-1}\tri.\label{eq:E_hru}
\end{align}
In the M-step, the parameters in $\Theta$ are updated as follows:
\begin{align}
    \rh &\leftarrow \frac{\hrh+\beta}{\alpha+2},\label{eq:M_rh}\\
    \li &\leftarrow \frac{1}{J}\sum_{j}\frac{1}{\tru}\vpi^{\HH}\hru\vpi,\label{eq:M_li}\\
    \ri &\leftarrow \rit+\li\bi\bi^{\HH},\\
    \ru &\leftarrow \frac{1}{M}\tr((\ri)^{-1}\hru).\label{eq:M_ru}
\end{align}

\subsubsection{Initialization of parameters\label{sect:init}}
We employ ILRMA estimates to initialize the variances $\rh$ and $\ru$ as
\begin{align}
    \rh &= \sum_{k}t_{ik,n_h}v_{kj,n_h},\label{eq:rhinit}\\
    \ru &= \frac{1}{M}(\hat{\bm{y}}_{ij}^{(u)})^{\HH}(\rit)^{+}\hat{\bm{y}}_{ij}^{(u)},\label{eq:ruinit}
\end{align}
where $t_{ik,n_h}$ and $v_{kj,n_h}$ are the NMF parameters in the low-rank source model obtained by ILRMA, 
$k=1,\dots,K$ is the index of the NMF bases, $K$ is the number of NMF bases,
$^{+}$ denotes the pseudoinverse, and $\hat{\bm{y}}_{ij}^{(u)}$ is the scale-fixed source image of
diffuse noise calculated as 
\begin{align}
    \hat{\bm{y}}_{ij}^{(u)} &=\rbm{W}_{i}^{-1}(\bm{w}_{i,1}^{\HH}\xij,\dots,\bm{w}_{i,n_h-1}^{\HH}\xij,0,\nonumber\\
    &\phantom{=}\bm{w}_{i,n_h+1}^{\HH}\xij,\dots,\bm{w}_{i,M}^{\HH}\xij)^{\top}.
\end{align}
The parameter $\li$ is initialized by the minimum nonzero eigenvalue of $\rit$.

\section{Proposed Method}
\subsection{Motivation\label{sec:vital}}
Update rules (\ref{eq:E_hrh}) and (\ref{eq:E_hru}) involve an inverse matrix operation of $M\times M$ matrices at each time-frequency slot.
Thus, their computational complexity is $O(IJM^3)$.
Such a heavy computational load restricts their implementation on low-resource hardware, such as hearing-aid devices.
To avoid this problem, we propose an efficient update algorithm
that greatly accelerates the estimation of parameters by expanding matrix inversion.
The acceleration consists of two steps:
(i) expanding matrix inversion using the Sherman--Morrison formula
and (ii) expanding matrix inversion using the pseudoinverse of matrices.
\subsection{Key Concept for Acceleration}
\subsubsection{First-stage acceleration}
Using the Sherman--Morrison formula, we can expand the inverse of matrix (\ref{eq:rijx}) as follows:
\begin{align}
    &\phantom{=}(\trijx)^{-1}\nonumber\\
    &= \frac{1}{\tru}(\tri)^{-1}-\frac{\frac{\trh}{(\tru)^2}(\tri)^{-1}\ai(\ai)^{\HH}(\tri)^{-1}}{1+\frac{\trh}{\tru}(\ai)^{\HH}(\tri)^{-1}\ai}\\
    &= \frac{1}{\tru}\left((\tri)^{-1}-\frac{\trh}{\tru+\trh(\ai)^{\HH}(\tri)^{-1}\ai}\right.\nonumber\\
    &\phantom{=}\cdot(\tri)^{-1}\ai(\ai)^{\HH}(\tri)^{-1}\Biggr).\label{eq:rijxinv}
\end{align}
Note that $\tri=\rit+\tli\bi\bi^{\HH}$ is invertible.
This expansion enables us to reduce the computational complexity
of (\ref{eq:E_hrh}) and (\ref{eq:E_hru}) from $O(IJM^3)$ to $O(IM^3+IJM^2)$, where the two terms respectively correspond to
matrix inversion at each frequency bin and multiplication of a matrix and vector at each time-frequency slot.

\subsubsection{Second-stage acceleration}
Using $\rit\bi=\bm{0}$ and $\|\bi\|_2=1$, we can expand the inversion $(\ri)^{-1}=(\rit+\li\bi\bi^{\HH})^{-1}$ using the pseudoinverse of $\rit$ as
\begin{align}
    (\ri)^{-1} &= (\rit)^{+}+\frac{1}{\li}\bi\bi^{\HH}.\label{eq:riinv}
\end{align}
Since $\rit$ is fixed in rank-constrained SCM estimation,
neither matrix inversion nor pseudoinversion is necessary in the parameter update step
if the pseudoinverse $(\rit)^{+}$ is calculated in advance.
Hence, the number of matrix inversions is reduced from $I$ to zero with expression (\ref{eq:riinv}).
Furthermore, no multiplication involving a matrix or vector is required, as shown in Sect.~\ref{sec:accel2upd}.

\subsection{Accelerated Update Rule\label{sec:fastupdate}}
\subsubsection{First-stage acceleration}
For the sake of simplicity, we define the following variables:
\begin{align}
    \raa&:=(\ai)^{\HH}(\ri)^{-1}\ai,\label{eq:raa}\\
    \traa&:=(\ai)^{\HH}(\tri)^{-1}\ai,\label{eq:traa}\\
    \rax&:=(\ai)^{\HH}(\ri)^{-1}\xij,\label{eq:rax}\\
    \trax&:=(\ai)^{\HH}(\tri)^{-1}\xij,\label{eq:trax}\\
    \rxx&:=\xij^{\HH}(\ri)^{-1}\xij,\label{eq:rxx}\\
    \sap &:= (\ai)^{\HH}\vpi,\label{eq:sap}\\
    \spx &:= \vpi^{\HH}\xij,\label{eq:spx}\\
    \tal&:=\frac{\trh}{\tru+\trh\traa}.\label{eq:tal}
\end{align}
From (\ref{eq:rijxinv}), we obtain
\begin{align}
    (\ai)^{\HH}(\trijx)^{-1}\ai&= \frac{1}{\tru}\left(\traa-\tal(\traa)^2\right)\\
    &= \frac{\traa}{\tru+\trh\traa},\\
    (\ai)^{\HH}(\trijx)^{-1}\xij &= \frac{\trax}{\tru+\trh\traa}.
\end{align}
The update rule (\ref{eq:E_hrh}) in the E-step can be rewritten as
\begin{align}
    \hrh &= \tal(\tru+\tal|\trax|^2).
\end{align}
Hence, we update $\rh$ in the M-step as
\begin{align}
    \rh &\leftarrow \frac{\tal\left(\tru+\tal|\trax|^2\right)+\beta}{\alpha+2}.\label{eq:1st_rh}
\end{align}
In addition, for update rule (\ref{eq:M_li}) in the M-step,
we can utilize the following equations, which are obtained using (\ref{eq:rijxinv}):
\begin{align}
    &\phantom{=}\tri(\trijx)^{-1}\nonumber\\
    &= \frac{1}{\tru}\left(\rbm{I}-\tal\ai(\ai)^{\HH}(\tri)^{-1}\right),\label{eq:ririjxinv}\\
    &\phantom{=}\vpi^{\HH}\tri(\trijx)^{-1}\tri\vpi\nonumber\\
    &= \frac{1}{\tru}\left(\tli-\tal|\sap|^2\right),\label{eq:bririjxrib}\\
    &\phantom{=}\vpi^{\HH}\tri(\trijx)^{-1}\xij\nonumber\\
    &= \frac{1}{\tru}\left(\spx-\tal\overline{\sap}\trax\right),\label{eq:bririjxx}
\end{align}
where $\overline{\cdot}$ denotes the complex conjugate.
Using (\ref{eq:ririjxinv})--(\ref{eq:bririjxx}), we obtain
\begin{align}
    \li &\leftarrow \frac{1}{J}\sum_{j}\left(\tal|\sap|^2 +\frac{1}{\tru}\left|\spx-\tal\overline{\sap}\trax\right|^2\right).\label{eq:1st_li}
\end{align}
Moreover, for update rule (\ref{eq:M_ru}) in the M-step,
the following equation can be used to derive an efficient update rule:
\begin{align}
    \hru &= \tal(\tru+\tal|\trax|^2)\ai(\ai)^{\HH}+\xij\xij^{\HH}\nonumber\\
    &\phantom{=}\mbox{}-\tal\trax\ai\xij^{\HH}-\tal\overline{\trax}\xij(\ai)^{\HH}.
\end{align}
The parameter $\ru$ can be updated as
\begin{align}
    \ru &\leftarrow \frac{1}{M}\left(\tal\raa(\tru+\tal|\trax|^2)\right.\nonumber\\
    &\phantom{\leftarrow}\mbox{}+\rxx-2\tal\Re[\trax\overline{\rax}]\Bigr).\label{eq:1st_ru}
\end{align}
\begin{algorithm}[t]
    Run ILRMA and calculate $\ai$, $\rit$, and $\bi$\;
    Initialize $\rh$, $\ru$ and $\li$ as described in Sect.~\ref{sect:init}\;
    Calculate $\sap$ and $\spx$ by (\ref{eq:sap}) and (\ref{eq:spx}) for all $i$ and $j$\;
    Calculate $\raa$ and $\rax$ by (\ref{eq:raa}) and (\ref{eq:rax}) for all $i$ and $j$\;
    \Repeat{converge}{
        $\trh\leftarrow\rh$ for all $i$ and $j$\;
        $\tru\leftarrow\ru$ for all $i$ and $j$\;
        $\tli\leftarrow\li$ for all $i$\;
        $\traa\leftarrow\raa$ for all $i$\;
        $\trax\leftarrow\rax$ for all $i$ and $j$\;
        Calculate $\tal$ by (\ref{eq:tal}) for all $i$ and $j$\;
        Update $\rh$ by (\ref{eq:1st_rh}) for all $i$ and $j$\;
        Update $\li$ by (\ref{eq:1st_li}) for all $i$\;
        Calculate $\raa$, $\rax$, and $\rxx$ by (\ref{eq:raa}), (\ref{eq:rax}), and (\ref{eq:rxx}) for all $i$ and $j$\;
        Update $\ru$ by (\ref{eq:1st_ru}) for all $i$ and $j$\;
    }
\caption{Algorithm for the first-stage acceleration update rule of the rank-constrained SCM estimation\label{alg:1st}}
\end{algorithm}
The update algorithm for the first-stage acceleration is summarized in Algorithm \ref{alg:1st}.
Note that we must update $\raa$, $\rax$, and $\rxx$ using up-to-date $\ri$.
In summary, in the first-stage acceleration, only the matrix inversions
of $\ri$ and $\tri$ are necessary for each frequency bin,
requiring a calculation cost of $O(IM^3)$.

\subsubsection{Second-stage acceleration\label{sec:accel2upd}}
On the basis of (\ref{eq:riinv}), we can further accelerate the update algorithm.
We define the quadratic terms
\begin{align}
    \taa &:= (\ai)^{\HH}(\rit)^{+}\ai,\label{eq:taa}\\
    \tax &:= (\ai)^{\HH}(\rit)^{+}\xij,\label{eq:tax}\\
    \txx &:= \xij^{\HH}(\rit)^{+}\xij.\label{eq:txx}
\end{align}
These terms do not depend on the variables $\rh$, $\ru$, and $\li$ and can be calculated before the update iteration.
Matrix inversions appearing in $\raa$, $\traa$, $\rax$, $\trax$, and $\rxx$
can be transformed using (\ref{eq:riinv}) as follows:
\begin{align}
    \raa &= \taa+\frac{|\sap|^2}{\li},\\
    \traa &= \taa+\frac{|\sap|^2}{\tli},\label{eq:traa_accel}\\
    \rax &= \tax+\frac{\sap\spx}{\li},\\
    \trax &= \tax+\frac{\sap\spx}{\tli},\\
    \rxx &= \txx+\frac{|\spx|^2}{\li}.
\end{align}
Thus, the update rules with the second-stage acceleration consist of 
only scalar operations and can be obtained as follows:
\begin{align}
    \rh &\leftarrow \frac{\tal\left(\tru+\tal\left|\tax+\displaystyle\frac{\sap\spx}{\tli}\right|^2\right)+\beta}{\alpha+2},\\
    \li &\leftarrow \frac{1}{J}\sum_j\Biggl(\tal|\sap|^2\nonumber\\
    &\phantom{\leftarrow}\left.+\frac{1}{\tru}\left|\spx-\tal\overline{\sap}\left(\tax+\frac{\sap\spx}{\tli}\right)\right|^2\right),\\
    \ru &\leftarrow \frac{1}{M}\Biggl(\tal\left(\taa+\frac{|\sap|^2}{\li}\right)\nonumber\\
    &\phantom{\leftarrow}\cdot\left(\tru+\tal\left|\tax+\frac{\sap\spx}{\tli}\right|^2\right)+\txx\nonumber\\
    &\phantom{\leftarrow}+\frac{|\spx|^2}{\li} -2\tal\Re\left[\left(\tax+\frac{\sap\spx}{\tli}\right)\right.\nonumber\\
    &\phantom{\leftarrow}\left.\left.\cdot\left(\overline{\tax}+\frac{\overline{\sap}\overline{\spx}}{\li}\right)\right]\right).
\end{align}
The update algorithm for the second-stage acceleration is summarized in Algorithm \ref{alg:2nd}.
\begin{algorithm}[t]
    Run ILRMA and calculate $\ai$, $\rit$, and $\bi$\;
    Initialize $\rh$, $\ru$ and $\li$ as described in Sect.~\ref{sect:init}\;
    Calculate $\sap$ and $\spx$ by (\ref{eq:sap}) and (\ref{eq:spx}) for all $i$ and $j$\;
    Calculate $\taa$, $\tax$, and $\txx$ by (\ref{eq:taa}), (\ref{eq:tax}), and (\ref{eq:txx}) for all $i$ and $j$\;
    \Repeat{converge}{
        $\trh\leftarrow\rh$ for all $i$ and $j$\;
        $\tru\leftarrow\ru$ for all $i$ and $j$\;
        $\tli\leftarrow\li$ for all $i$\;
        Calculate $\tal$ by (\ref{eq:tal}) and (\ref{eq:traa_accel}) for all $i$ and $j$\;
        Update $\rh$ by (\ref{eq:1st_rh}) for all $i$ and $j$\;
        Update $\li$ by (\ref{eq:1st_li}) for all $i$\;
        Update $\ru$ by (\ref{eq:1st_ru}) for all $i$ and $j$\;
    }
\caption{Algorithm for second-stage acceleration update rule of the rank-constrained SCM estimation\label{alg:2nd}}
\end{algorithm}

\subsection{Advantage of Proposed Accelerated Update Rules}
In general, the complexity of size-$M$ matrix inversions and matrix multiplications is $O(M^3)$,
and that of multiplications of a matrix and a vector is $O(M^2)$.
The iteration-wise computational complexities of the naive update rule 
and the proposed update rules with the first-stage and second-stage accelerations are summarized in Table \ref{tbl:complexity}.
Note that initialization with (\ref{eq:rhinit}) and (\ref{eq:ruinit}) is required for all the methods.
The proposed algorithms can reduce the complexity via the use of (\ref{eq:rijxinv}) and (\ref{eq:riinv}).
In particular, the second-stage acceleration has greatly improved efficiency
because its computational cost does not depend on the number of microphones $M$,
thus enabling us to apply BSS with a large-scale microphone array.
\begin{table}[t]
    \centering
    \caption{Computational complexity of initialization\protect\linebreak and iterative update for each method\label{tbl:complexity}}
    \begin{tabular}{ccc}\bhline{1.5pt}
        Method&Initialization&Iterative update\\\hline
        Naive & $O(IM^3+IJM^2)$ & $O(IJM^3)$\\
        First-stage accel. & $O(IM^3+IJM^2)$ & $O(IM^3+IJM^2)$\\
        Second-stage accel. & $O(IM^3+IJM^2)$ & $O(IJ)$\\\bhline{1.5pt}
    \end{tabular}
\end{table}

For example, $I=513$, $J=275$, and $M=4$ for the conditions described in Sect.~\ref{sect:exp_cond}.
In such a case, the naive update rule requires $IJ=141075$ inverse matrix operations per iteration,
which is no longer necessary for the second-stage acceleration.

\section{Experiments}
\subsection{Experimental Conditions\label{sect:exp_cond}}
\begin{figure}[t]
    \centering
    \includegraphics[width=0.65\linewidth,pagebox=mediabox]{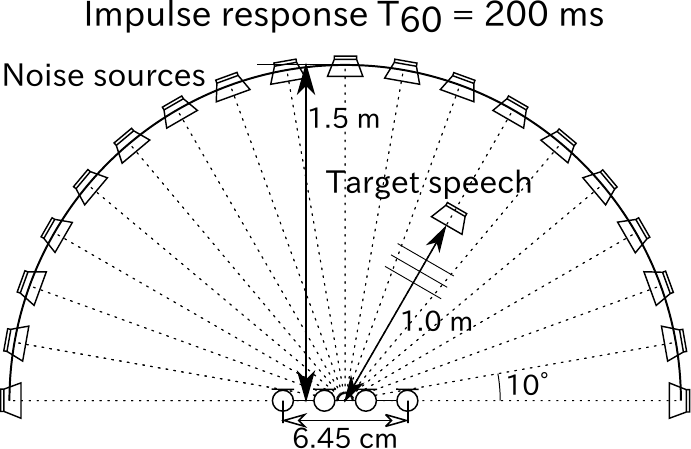}
        \caption{\label{fig:reccond}Recording conditions of impulse responses.}
\end{figure}

To compare the efficacy of the proposed algorithms and the separation quality,
we conducted an audio BSS experiment with simulated mixtures of
a directional target speech and diffuse babble noise.
We compared three methods:
FastMNMF~\cite{Sekiguchi2019_FastMNMF},
FastMNMF initialized by ILRMA (ILRMA+FastMNMF),
and rank-constrained SCM estimation initialized by ILRMA~\cite{Kubo2019_RCSCM}.
For rank-constrained SCM estimation, we compared three update algorithms, namely,
the naive update rule (\textit{Naive}) and the proposed update rules 
with the first and second accelerations (\textit{Proposed 1st-stage accel.} and \textit{Proposed 2nd-stage accel.}, respectively).
In ILRMA, the observation $\xij$ was preprocessed via a sphering transformation using principal component analysis.
For ILRMA and FastMNMF, all the NMF variables were initialized by nonnegative random values.
The demixing matrix $\rbm{W}_{i}$ in ILRMA and the spatial covariance matrix in FastMNMF were initialized by the identity matrix.
For ILRMA+FastMNMF, the NMF variables were initialized using the estimates of ILRMA.
Also, the spatial covariance matrix was initialized using $\bm{a}_{i,n}\bm{a}_{i,n}^{\HH}+\varepsilon\sum_{n'\neq n}\bm{a}_{i,n'}\bm{a}_{i,n'}^{\HH}$ for ILRMA+FastMNMF, where $\bm{a}_{i,n}$ was estimated by ILRMA and $\varepsilon$ was set to $10^{-5}$.
These methods were implemented in MATLAB (R2019a),
and the computation was performed on an Intel Core i9-7900X (3.30 GHz, 10 cores) CPU.
The dry sources of the directional target speech and diffuse babble noise
were obtained from the JNAS speech corpus~\cite{KItou1999_JNAS}.
They were convoluted with the impulse responses shown in Fig.~\ref{fig:reccond} to simulate a mixture of 8.7~s length.
The target source was located 30$^{\circ}$ clockwise from the normal to a microphone array,
the 19 loudspeakers used to simulate diffuse noise were arranged at intervals of 10$^{\circ}$, 
the size of the recording room for these impulse responses was 3.9 m $\times$ 3.9 m,
and its reverberation time was about 200 ms.
Note that the diffuse babble noise was produced by convoluting 19 independent speeches with each impulse response.
The speech-to-noise ratio was set to 0 dB.
The other conditions are shown in Table~\ref{tbl:expcond}.

\subsection{Results}
\begin{table}
    \centering
    \caption{\label{tbl:expcond}Experimental conditions}
    \begin{tabular}{cc}\bhline{1.5pt}
    Sampling frequency & 16 kHz\\\hline
    \multirow{2}{*}{STFT} & 64-ms-long Hamming\\
    & window with 32 ms shift\\\hline
    Number of NMF bases $K$ & 10 for each source\\\hline
    Number of iterations &\multirow{2}{*}{50}\\
    in ILRMA & \\\hline
    Maximum number of iterations & \multirow{2}{*}{200}\\
    in methods except ILRMA\\\hline
    Shape and scale parameters & $\alpha=1.1$ and $\beta=10^{-16}$\\\bhline{1.5pt}
    \end{tabular}
\end{table}
\begin{figure}[t]
    \centering\includegraphics[width=\linewidth,pagebox=cropbox]{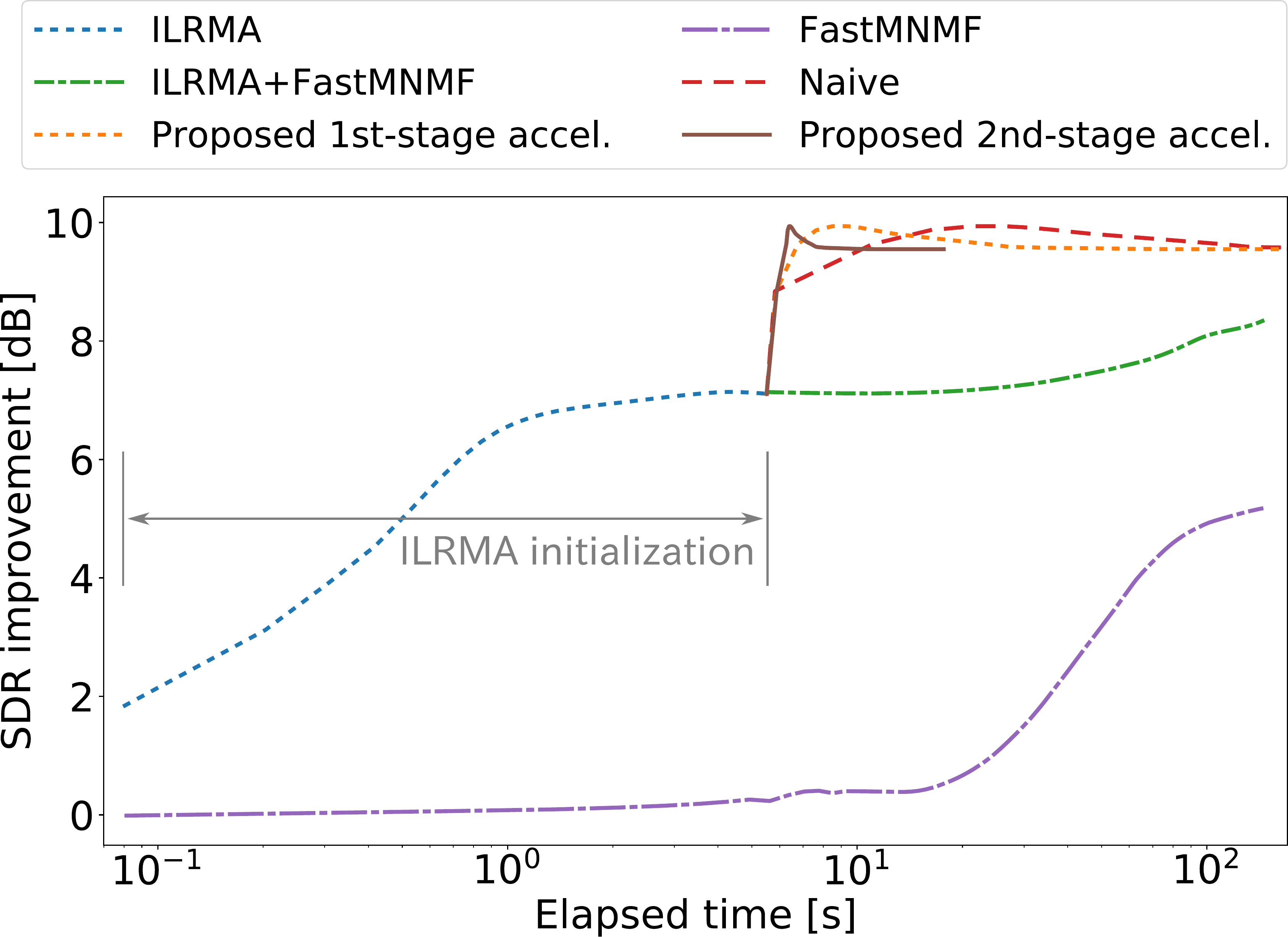}
    \caption{\label{fig:SDR_elapse} SDR behaviors with respect to elapsed time.}
\end{figure}
\begin{figure}[t]
    \centering\includegraphics[width=\linewidth,pagebox=artbox]{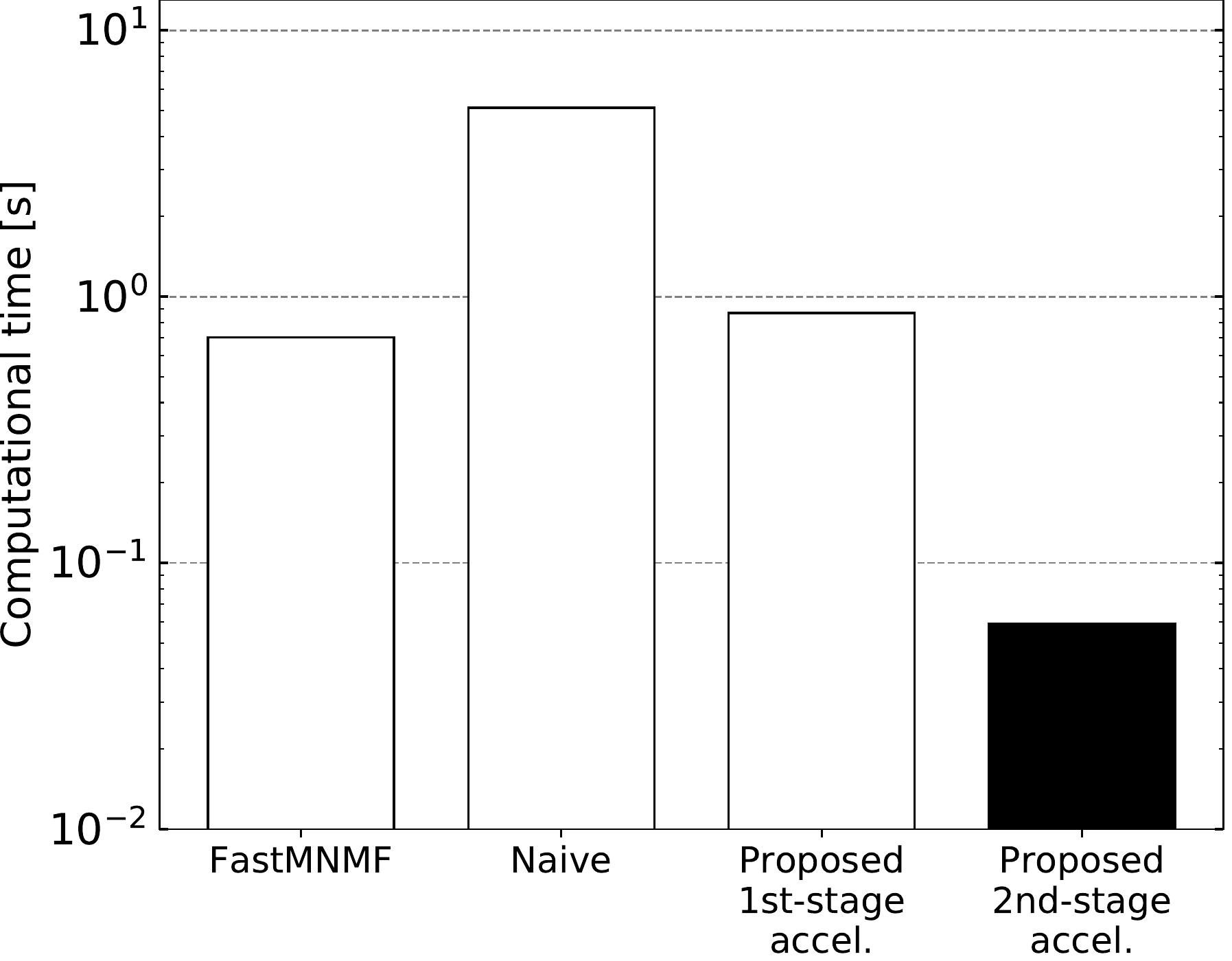}
    \caption{\label{fig:elapse} Average computational time of one iteration for each method.}
\end{figure}
We evaluated each method using the source-to-distortion ratio (SDR)~\cite{EVincent2006_BSSEval}
of the directional target speech, which measures the separation quality and the absence of distortion.
For each of the methods, the SDR behaviors with respect to the elapsed time are shown in Fig.~\ref{fig:SDR_elapse}.
Note that all methods except for FastMNMF are initialized using ILRMA.
Multichannel Wiener filtering using ILRMA estimates improves the SDR value to some extent,
and FastMNMF or rank-constrained SCM estimation further improves the separation quality.
From Fig.~\ref{fig:SDR_elapse}, we can confirm that
the second-stage acceleration achieves the fastest and most efficient target speech extraction.

Fig.~\ref{fig:elapse} shows the average computational time of one iteration for each method.
The proposed algorithm with the second-stage acceleration was 87 times faster
than the naive update rule and 12 times faster than FastMNMF with the four-microphone condition.

\section{Conclusion}
In this paper, we presented new accelerated update rules for rank-constrained SCM estimation,
enabling effective extraction of a directional target source contaminated by diffuse noise.
We derived update rules by expanding matrix inversion in the naive update rule.
The experiment showed that the proposed method achieved an 87 times faster update
than the naive update rule and a 12 times faster update than FastMNMF.

\section*{Acknowledgments}
This work was partly supported by SECOM Science and Technology Foundation and JSPS KAKENHI Grant Numbers 17H06101, 19H01116, and 19K20306.

\bibliographystyle{IEEEbib}
\bibliography{refs}
\end{document}